\documentclass[journal, onecolumn]{IEEEtran}
\usepackage{cite}
\usepackage{amsmath}
\usepackage{graphicx}
\usepackage{algorithm}
\usepackage{algpseudocode}
\usepackage{multirow}
\usepackage[table,xcdraw]{xcolor}
\usepackage{authblk}
\usepackage[bookmarks=false]{hyperref}
\usepackage{xcolor}

\algblock{Input}{EndInput}
\algnotext{EndInput}
\algblock{Output}{EndOutput}
\algnotext{EndOutput}

\newcommand*{\affaddr}[1]{#1} % No op here. Customize it for different styles.
\newcommand*{\affmark}[1][*]{\textsuperscript{#1}}
\newcommand*{\email}[1]{\textit{#1}}

\begin{document}

%\title{Email Embeddings for Phishing Detection}
% or
\title{Phishing Detection through Email Embeddings}

% single blind review
\author{%
Luis Felipe Gutiérrez\affmark[1], Faranak Abri\affmark[1], Miriam Armstrong\affmark[2], Akbar Siami Namin\affmark[1], and Keith S. Jones\affmark[2]\\
\affaddr{\affmark[1]Department of Computer Science,}
\affaddr{\affmark[2]Department of Psychological Sciences}\\
\affaddr{\affmark[]Texas Tech University}\\
%\affaddr{\affmark[]Lubbock, TX, USA}\\
\email{\{Luis.Gutierrez-Espinoza, faranak.abri, miriam.armstrong, akbar.namin, keith.s.jones\}@ttu.edu}\\
%\affaddr{\LaTeX\ University}%
}

\IEEEoverridecommandlockouts
\IEEEpubid{\makebox[\columnwidth]{978-1-7281-6251-5/20/\$31.00~\copyright2020 IEEE \hfill} \hspace{\columnsep}\makebox[\columnwidth]{ }}

\maketitle
%IEEEpubidadjcol

\begin{abstract}

The problem of detecting phishing emails through machine learning techniques has been discussed extensively in the literature. Conventional and state-of-the-art machine learning algorithms have demonstrated the possibility of building classifiers with high accuracy. The existing research studies treat phishing and genuine emails through general indicators %. as bulks 
and thus it is not exactly clear what phishing features are contributing to variations of the classifiers. In this paper, we crafted a set of phishing and legitimate emails with similar indicators in order to investigate whether these cues are captured or disregarded by email embeddings, i.e., vectorizations. We then fed machine learning classifiers with the carefully crafted emails to find out about the performance of email embeddings developed. Our results show that using these indicators, email embeddings techniques is effective for classifying emails as phishing or legitimate\footnote{This paper is the pre-print of a paper to appear in the proceedings of the IEEE International Conference on BigData (Workshop - BigData'20) entitled: ``{\it Email Embeddings for Phishing Detection}.''}. 

\end{abstract}

\begin{IEEEkeywords}
    Natural Language Processing, Phishing Emails, Email Embeddings 
\end{IEEEkeywords}

\section{Introduction}

Autonomous devices are the integral part of cyber-Physical Systems (CPS) \cite{DBLP:conf/bigdataconf/ChatterjeeND18}. These devices are powered by elegant software applications and utilities to protect against adversarial attacks. Information theoretical solutions along with AI-enabled autonomous agents are capable of protecting these critical components from cyber attacks. However, in addition to these hardware and software devices, there is another autonomous entity in CPS that is harder to harden their security defense systems, i.e., humans. 

It is known that humans are the weakest link in the information security chain. Adversarial attacks often exploit this weakness through sending malicious links to individuals who are operating critical infrastructure with the hope that the operators visit malicious links and Websites. These types of phishing attacks can be in the form of emails, adware, or even malicious fake Websites \cite{DBLP:conf/icmla/Abri2020}. 

In spite of advancement in security technologies and controls, cyber security attacks through phishing still remain the number one choice of attackers due to their higher success rates. Even though many antivirus software and blocking strategies are able to detect spamicity of emails, these malicious emails are able to get the attention of the target receivers, pass through the spam detection tools, and thus being activated. 

Machine learning-based approaches have revolutionized the detection mechanism of spam and phishing emails. However, the goodness and accuracy of the detection power of these learning-based algorithms heavily rely on the type of historical data, as known as training data. Existing studies confirm that phishing emails are distinguishable from genuine ones based on features. However, it is unclear whether the historical data are rich enough to train the classifiers properly. More specifically, an important question is whether the machine learning detection-based models are content-aware or they are content agnostic.   

To address this grand problem, this paper studies the performance of email embeddings (i.e., email vectorization) to detect phishing emails. %To do so, we have carefully created a set of phishing and genuine emails with similar contents except that some features are manually changed in order to superimpose some cues into the content of emails. More specifically, we manipulated three {\it sentimental}  factors: 1) the presence or absence of language about {\it urgency}, 2) the {\it amount} the victim would purportedly {\it gain} or {\it lose}, and 3) whether the victim would purportedly gain or lose. %what is called the``{\it persuasion principals}''. Examples of such principles include the use of authorities, Scarcity, and liking. 
We created 12 genuine and 12 phishing emails. %in which the above three factors are controlled. 
The systematically crafted emails are then vectorized and fed into well-known machine learning classifiers. The results demonstrate that, even though the contents of both phishing and legitimate emails are similar, the email embeddings technique is able to distinguish between phishing and genuine emails. To capture the content of the emails into account, we then embed emails using doc2vec, a vectorization approach to capture the semantic of a given text. %, even in the presence of similar contents of both types of emails. 

In our previous work \cite{DBLP:conf/compsac/Gutierrez-Espinoza20,
DBLP:conf/icmla/Abri2020}, we studied the usage of linguistic features in the context of fake reviews. We observed that it is possible to detect fake reviews using linguistic features. In this paper, we are interested in exploring whether it is possible to detect phishing emails when the contents of both phishing and legitimate emails are somewhat similar. %More specifically, we would like to investigate the effectiveness of linguistic approaches to detect phishing emails in the presence of the above sentimental factors. 
The key contributions of this paper are as follows: 

\begin{itemize}
    \item[--] We introduce a carefully and systematically crafted set of phishing and legitimate emails to support this line of research. 
    \item[--] The performance of email embeddings techniques is presented through a number of machine learning classifiers.
    \item[--] The results show that even in the presence of similar contents, the email embeddings are able to distinguish legitimate emails from the phishing ones. 
\end{itemize}

This paper is organized as follows: Section \ref{sec:relatedwork} reviews the related work in this line of research. In Section \ref{sec:classifiers}, a brief technical background of the machine learning models and embeddings is presented. Section \ref{sec:setup} presents the experimental setup and procedure. The results of the study are presented in Section \ref{sec:results}. Section \ref{sec:conclusion} concludes the paper and highlights future research directions.

\section{Related work}
\label{sec:relatedwork}

Machine learning techniques are broadly used for spam detection. The most important step for developing such a detection framework is extracting features that are fed into classifiers. A very common method used for this step is Bag of Words (BOW), which utilizes the occurrence frequency of the words or group of words. Although this method is computationally fast and easy to implement, it has several limitations such as ignoring semantics, word order, and a huge number of features that imposes the curse of dimensionality on machine learning classifiers~\cite{2019Heliyon}.

Term frequency-inverse document frequency (tf-idf) is another method for feature extraction that is commonly used in search engines and is useful for spam detection. It considers the frequency of only the key words by checking the occurrence frequency of a word in a text or document and comparing it to its occurrence in the whole document or corpus. Therefore, words that are repeated in small sections have higher scores compared to commonly used words in the whole content.

BOW or tf-idf are less efficient methods for detecting spams, such as phishing emails, which have more specific content properties such as a URL link to a malicious source. Given that sometimes it is desirable to pick out phishing emails especially from spam emails, more robust features are needed. Doc2Vec~\cite{le2014distributed} is a document embedding technique in which similar documents have similar encodings semantically. It is an unsupervised neural network,  which predicts the words in a document.

%------------
Duzi et al.~\cite{2020Duzi} proposed an ensemble method that detects spam emails by combining the classifying results obtained from two feature sets. These feature sets extracted using Doc2Vec and tf-idf methods. They compared different classifiers with these feature sets on two datasets: $33,702$ emails from the Enron dataset~\cite{2004enron} and $2,314$ emails from the Ling spam corpus. Using support vector machines, they achieved 98.27\% for accuracy and 98.97\% for f-score on the Ling spam dataset and 96.16\% for accuracy and 96.07\% for f-score on the Enron dataset, respectively.

%------------------
Akinyelu and Adewumi~\cite{2014Akinyelu} evaluated a random forest (RF) model for classifying phishing emails from legitimate ones. Using two datasets ~\cite{spamassassin2006} for legitimate emails and~\cite{Phishingcorpus2006} for phishing emails, a new dataset with a total of $2,000$ emails was made. They extracted 15 features such as "URLs Containing IP Address" and "Presence of Javascript" from samples. Using 10 fold cross validation by a RF classifier, the best accuracy achieved with 99.7\% along with 99.47\%, 97.5\%, 98.45\% for precision, recall and f-score, respectively. They also examined their model on datasets with different sizes and showed that the performance of the classifier increases using more email samples.
%-------------

Unnithan et al.~\cite{2018Unnithan} examined several classifiers for phishing email detection. They used the dataset shared by the "IWSPA-AP 2018" workshop, including train and test subsets for emails with headers and emails without headers separately. The subset of no-header emails contains $4,300$ samples for the test data and $5,700$ emails as train data, including $5,088$ legitimate and $612$ phishing emails. Two sets of features were generated using the Term frequency-inverse document frequency (tf-idf) and Doc2Vec techniques. They trained seven different models including: Decision Tree (DT), Naive Bayes (NB), Ada-boost, Logistic Regression (LR), K-nearest neighbour (KNN), Support vector machine (SVM), and Random Forest (RF). They applied the models to both subsets of the data (emails with and without headers) using both feature sets. Using the SVM classifier with Doc2Vec feature set on both datasets, they achieved the best results including: $3,825$ true positives (TP), 0 true negative (TN), $475$ false positive (FP), and $0$ false negative (FN) on emails without header.

\section{Models and Algorithms}
\label{sec:classifiers}

\subsection{Feature Extraction}

\subsubsection{Doc2Vec}

\begin{figure}[h]
  \centering
  \includegraphics[width=0.8\linewidth]{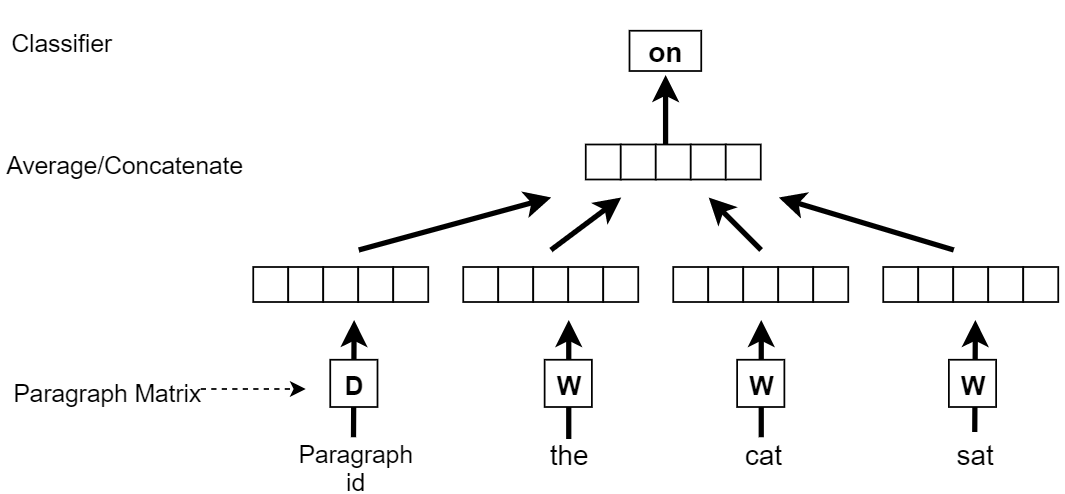}
  \caption{Doc2Vec model (adapted from \cite{le2014distributed}).}
  \label{fig:doc2vec-model}
      \vspace{-0.1in}
\end{figure}

Doc2Vec is a technique to generate document embeddings through a neural network-based approach \cite{le2014distributed}, similar to what Word2Vec \cite{mikolov2013efficient} achieves with words. Figure \ref{fig:doc2vec-model} shows the Doc2Vec model adapted from \cite{le2014distributed}. As it is observable, the inclusion of a {\it paragraph id} in the training phase is a difference with respect to the original Word2Vec model. Note that although in \cite{le2014distributed} the term used is {\it paragraph id}, normally, this method is used to embed full documents. Given that all words in a specific document are trained using the id assigned to the document, the resulting vector (i.e., embedding) for the document encodes meaning according to the words it contains. Furthermore, as both word and document embeddings are generated during the same training procedure, both words and documents are embedded in the same semantic vector space.

Hence, the resulting vector space has the capability to encode semantic similarities, i.e., vectors that are close together tend to share semantic properties, as Doc2Vec works under the Distributional Hypothesis of linguistics \cite{mikolov2013efficient} (i.e., words that are used in similar contexts tend to hold similar meaning).

Doc2Vec generates dense feature vectors, in contrast to the sparse representations produced by frequency-based techniques (e.g., BOW). Because of this, there exists a trade-off between sparsity of the features and interpretability, as the features extracted by Doc2Vec are considered opaque, since there is no direct interpretation for the values in each dimension. 

\subsection{Dimensionality Reduction}

\subsubsection{Principal Component Analysis}

Principal Component Analysis (PCA) is a technique that is commonly used in dimensionality reduction \cite{shlens2014tutorial}. The dimensionality reduction is performed by considering the $m$ principal components in the dataset, while retaining as much variance of the original dataset as possible. However, there are three important assumptions of the given data when PCA is utilized \cite{shlens2014tutorial}:

\begin{enumerate}
    \item Linearity of the data should be present,
    \item Large variances denote an important structure, and
    \item The principal components are orthogonal.
\end{enumerate}

In addition, when PCA is used for visualization purposes, the value for $m$ (i.e., the number of principal components) is 2 or 3. In this work, we set $m = 2$ to visualize the data and also to explore the results of classification using the 2-D projection of the document embeddings.

\subsubsection{Kernel Principal Component Analysis}

In some cases, the linear nature of the data that PCA takes as an assumption might not hold, and a non-linear structure could explain the underlying phenomena more accurately. Using this idea, PCA can still be applied to such datasets using a non-linear mapping that takes the samples in the original space, often called input space, to a higher dimensional space, also called feature space, and then performing PCA in the resulting feature space \cite{ghodsi2006dimensionality}. The mapping function is often referred to as \textit{Kernel}, and the full algorithm as \textit{Kernel} PCA. Like the case of SVMs, several types of kernels can be utilized to model different patterns in the data. In our work, we used the Gaussian or RBF kernel.

\subsection{Classifiers}

Brief descriptions of the classifiers that we experimentally compared are as follows:

\subsubsection{Random Forest}
The random forest classifier, also known as ensemble of decision trees, combines the prediction of several weak classifiers (i.e. decision trees) and produces a final classification through voting.  %\cite{breiman2001random}. 
The main hyperparameter of this model is the number of weak classifiers to combine. Due to the fact that decision trees are weak classifiers, their corresponding hyperparameters are also present in this model.
    
\subsubsection{Support Vector Machine}
Support Vector Machines (SVMs) determine a hyperplane that maximizes the margin between itself and the nearest samples of each data class. %\cite{gunn1998support}. 
Such nearest samples are known as support vectors. This construction can be performed in the original input space, or a feature space that is generated by applying a kernel function to each pair of samples, which allows the decision boundary to be non-linear. %\cite{gunn1998support}. 
In our work, we used the linear, Radial Basis Function (RBF), polynomial, and sigmoid kernels. An SVM converges towards non-linearly separable classes using the $C$ hyperparameter, where $C$ controls the number of samples that can be misclassified, and its optimal value is problem-dependent. Moreover, there exist additional hyperparameters for each kernel, such as the degree in the polynomial kernel, or the gamma term in the RBF kernel.% \cite{gunn1998support}.

\subsubsection{Logistic Regression}

In binary classification, logistic regression assigns the probability that a sample, denoted by the feature vector $x$, belongs to a class $C_1$ as \cite{dreiseitl2002logistic}:

    \vspace{-0.15in}
\[
    P(C_1 | x, \alpha) = \frac{1}{1 + \exp{(- \alpha \cdot x)}}
\]

\noindent where $\alpha$ is the set of parameters of the model. The probability $P(C_2 | x, \alpha)$ is calculated as $1 - P(C_1 | x, \alpha)$. Usually, the learning of the $\alpha$ parameter is conducted through the optimization of the cross-entropy loss function \cite{dreiseitl2002logistic}.

\subsubsection{Naive Bayes}

The Naive Bayes classifier performs classification using the strong assumption that all features in a vector $X = (X_1, ... , X_n)$ are statistically independent with respect to the class $C$. This is,

    \vspace{-0.15in}
\[
    P(X | C) = \prod_{i = 1}^{n} P(X_i | C).
\]

This model assigns a predicted class label $\hat{y}$ according to

    \vspace{-0.15in}
\[
    \hat{y} = \arg \max_{k} P(C_k) \prod_{i = 1}^{n} P(x_i | C_k)
\]

\noindent where $C_k$ is the $k$-th class.

Because the features of our dataset are generated using Doc2Vec document embeddings, we used the Gaussian Naive Bayes in order to estimate the conditional probabilities with a dataset of continuous features. Given a class $C_k$, a variance $\sigma^2_k$, a mean $\mu_k$ (both $\sigma^2_k$ and $\mu_k$ are estimated from the samples), and an observed value $\hat{x}$ for the feature, the probability $P(x = \hat{x} | C_k)$ can be estimated using 

    \vspace{-0.15in}
\[
    P(x = \hat{x} | C_k) = \frac{1}{\sqrt{2 \pi \sigma^2_k}} \exp{(-\frac{(\hat{x} - \mu_k)}{2 \sigma^2_k})}.
\]

Although this assumption is unnatural and it does not hold for most situations, the Naive Bayes classifier has been adopted successfully in tasks such as spam filtering, medical diagnosis, and text classification \cite{rish2001empirical}

\iffalse
Doc2Vec \cite{le2014distributed} is a neural network-based document embedding technique that aims to generate document vectors in a similar fashion that Word2Vec \cite{mikolov2013efficient} does with word vectors. Like Word2Vec, Doc2Vec works under the Distributional Hypothesis (i.e., words used in similar contexts share similar semantics) in order to set the objective function during the training phase of the model. As a consequence of this, the framework for training document vectors is very similar to that of word vectors.
\fi

\section{Experimental Setup}
\label{sec:setup}

\subsection{Scripts and Libraries}

We developed Python 3 scripts for running our experiments. We used the classifiers' implementation provided by the Scikit-learn library. %\cite{pedregosa2011scikit}. 
To obtain the Doc2Vec document embeddings, we used the Gensim library \cite{rehurek_lrec}. For stemming, we used the Porter algorithm provided by the NLTK library \cite{loper2002nltk}.

\subsection{Data Collection}

 Twenty-four email stimuli were created for an experiment with human subjects. The content of the emails was modeled off legitimate emails found in the authors' inboxes or online. For all emails, a sender address, subject line, email body, and URL were created. The body of all emails was 50-100 words long and contained a hyperlink. Because the email stimuli were created to later show to research participants, this method for creating emails and the characteristics of the email stimuli are similar to those used in other phishing experiments \cite{canfield2016quantifying, downs2007behavioral, parsons2015design}.
 
 Twelve of the 24 email stimuli were legitimate emails, and the other 12 emails were phishing. All legitimate emails shared two characteristics. First, the URL in the legitimate emails went to a real and safe web address. All legitimate email URLs began with HTTPS. Second, the body of legitimate emails contained targeted messaging. All email stimuli were addressed to a fictional university student. Email stimuli with targeted messaging referenced the student by name, referenced their position as a student, or referenced the city in which the student attended university.

The 12 phishing emails were designed to mimic non-targeted phishing emails. The non-targeted phishing emails differed from the other types of emails in terms of both their URL and email body. The URLs of the non-targeted phishing emails did not contain HTTPS and had at least two additional characteristics of suspicious links \cite{chatterjee2019detecting}: five or more dots in the domain, over 75 characters long, contained an IP address, or contained misspellings. The email body contained no targeted messaging; non-targeted phishing emails were designed to be emails to could be sent to anyone.

\iffalse
\begin{table}[h]
    \caption{Number of samples per class (binary).}
    \label{tab:count-binary}
    \centering
    \begin{tabular}{cc}
    \hline
    \multicolumn{1}{|c|}{\bf Class} &
      \multicolumn{1}{c|}{\bf Number of samples}\\ \hline
    \multicolumn{1}{|c|}{Phishing} & \multicolumn{1}{c|}{12}\\ \hline
    \multicolumn{1}{|c|}{Legitimate} & \multicolumn{1}{c|}{12}\\ \hline
    \end{tabular}
\end{table}
\fi

\iffalse
\begin{table}[h]
    \caption{Number of samples per class (fine grained phishing).}
    \label{tab:count-3class}
    \centering
    \begin{tabular}{cc}
    \hline
    \multicolumn{1}{|c|}{\bf Class} &
      \multicolumn{1}{c|}{\bf Number of samples}\\ \hline
    \multicolumn{1}{|c|}{Spear phishing} & \multicolumn{1}{c|}{12}\\ \hline
    \multicolumn{1}{|c|}{Regular phishing} & \multicolumn{1}{c|}{12}\\ \hline
    \multicolumn{1}{|c|}{Legitimate} & \multicolumn{1}{c|}{12}\\ \hline
    
    \end{tabular}
\end{table}
\fi

\subsection{Data Preprocessing}

We used standard techniques for text preprocessing in order to clean the dataset. These are removal of 1) punctuation marks, 2) URLs, 3) email addresses, and 4) stopwords. We also used the Porter stemmer \cite{porter1980algorithm} to perform stemming for each remaining token. 

\iffalse

\begin{enumerate}
    \item Removal of punctuation marks.
    \item Remove URLs.
    \item Remove email addresses.
    \item Remove stopwords.
    \item Use the Porter stemmer to perform stemming for each remaining token.
\end{enumerate}
\fi

\subsection{Classification Metrics}

Consider a confusion matrix with counts of samples classified as True Positive (TP), True Negative (TN), False Positive (FP), and False Negative (FN). We report the results of our experiments using accuracy, $F_1$ measure, precision, and recall, which are defined as

    \vspace{-0.2in}
\begin{gather*}
    Accuracy = \frac{TP + TN}{TP + TN + FP + FN}, \\
    Precision = \frac{TP}{TP + FP}, \\
   Recall = \frac{TP}{TP + FN}, \\
    F_1 = \frac{2 \times Precision \times Recall}{Precision + Recall}.
\end{gather*}

%Due to the imbalanced nature of the 2-class dataset, we report the macro-weighted averaged $F_1$ scores to overcome class dominance. This metric is calculated as the average $F_1$ score for each label, weighted by the actual number of samples per label.

\subsection{Hyperparameters Tuning}

We performed several exhaustive grid search in order to find the best set of hyperparameters for SVM, Logistic Regression, and Random Forest. For SVM, we tried different values for the $C$ constant, kernel type, degree of the polynomial in case of using polynomial kernel, and the gamma value in case of the Gaussian kernel. For logistic regression, we tried different values for the regularization parameter. For random forests, we tried different number of estimators, maximum tree depth, minimum number of samples per leaf, minimum number of samples per split, and criterion to choose one split over others.

\subsection{Experiments Flowchart}

\begin{figure}[h]
  \centering
  \includegraphics[width=0.65\linewidth]{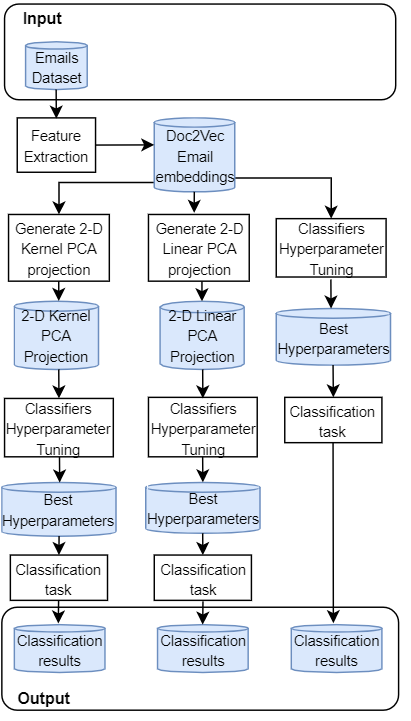}
  \caption{Flowchart of the experiments conducted in our work.}
  \label{fig:flowchart}
      \vspace{-0.2in}
\end{figure}

Figure \ref{fig:flowchart} shows the flowchart of the experiments carried out in our work. After the feature extraction step, there are three scenarios in which the classification task was applied: 1) directly using the 20-dimensional Doc2Vec email embeddings, 2) using the linear PCA 2-D projections of the embeddings, and 3) using the Kernel PCA 2-D projections of the embeddings. Whether it is using the full high-dimensional features or a 2-D projection of it, the next step is the hyperparameter tuning, in which the best hyperparameters for the classifiers are set. Once the hyperparameters are found, the classifiers are fit using them and report the classification metrics using 10-fold cross-validation.

\section{Results and discussion}
\label{sec:results}

\subsection{Classification}

We report the averages of accuracy and $F_1$ score obtained after running the experiments using 10-fold cross-validation. Table \ref{tab:fulldim-results} shows the results for the classifiers using the full 20-dimensional space provided by the document embeddings. SVM reports accuracy and $F_1$ score of 81.6\% and 76.6\%, respectively, which is significantly higher than that of the next classifier, the random forest.

\begin{table}[h]
    \caption{Performance scores for the classifiers in the 20-dimensional Doc2Vec feature space.}
    \label{tab:fulldim-results}
    \centering
    \begin{tabular}{ccccc}
    \hline
    \multicolumn{1}{|c|}{\bf Classifier} &
      \multicolumn{1}{c|}{\bf Accuracy} &
      \multicolumn{1}{c|}{\bf $F_1$ score} &
      \multicolumn{1}{c|}{\bf Precision} & 
      \multicolumn{1}{c|}{\bf Recall}  
      \\ \hline
    \multicolumn{1}{|c|}{SVM} & \multicolumn{1}{c|}{{\bf 0.816}} & \multicolumn{1}{c|}{{\bf 0.766}} &
    \multicolumn{1}{c|}{ \bf{0.750}} &
    \multicolumn{1}{c|}{ \bf{0.750}} \\ \hline
    \multicolumn{1}{|c|}{Logistic Regression} & \multicolumn{1}{c|}{0.616} &
    \multicolumn{1}{c|}{0.566} &
    \multicolumn{1}{c|}{ 0.600} &
    \multicolumn{1}{c|}{ 0.700} \\ \hline
    \multicolumn{1}{|c|}{Random Forest}           & \multicolumn{1}{c|}{0.583} & \multicolumn{1}{c|}{0.483} &
    \multicolumn{1}{c|}{0.533} &
    \multicolumn{1}{c|}{0.600} \\ \hline
    \multicolumn{1}{|c|}{Naive Bayes}          & \multicolumn{1}{c|}{0.600} & \multicolumn{1}{c|}{0.553} &
    \multicolumn{1}{c|}{0.616} &
    \multicolumn{1}{c|}{\bf{0.750}}\\ \hline
    \end{tabular}
        \vspace{-0.1in}
\end{table}

Table \ref{tab:linearpca-results} shows the results for the classifiers trained using only the first two components projections of linear PCA. In general, these results are consistently higher than those of in Table \ref{tab:fulldim-results}. This is an interesting finding as it suggests that the variance present in the first two components of PCA is not only enough to maintain the performance of the classifiers, but it also increases the performance. Random forest reports the highest accuracy and $F_1$ score with 91.6\% and 90.0\%, respectively. However the margin between the performance of random forest and the rest of the classifiers is also reduced, the SVM and Naive Bayes classifiers having similar values performance-wise.

\begin{table}[h]
    \caption{Performance scores for the classifiers in the 2-D linear PCA space.}
    \label{tab:linearpca-results}
    \centering
    \begin{tabular}{ccccc}
    \hline
    \multicolumn{1}{|c|}{\bf Classifier} &
      \multicolumn{1}{c|}{\bf Accuracy} &
      \multicolumn{1}{c|}{\bf $F_1$ score} &
      \multicolumn{1}{c|}{\bf Precision} &
      \multicolumn{1}{c|}{\bf Recall} \\ \hline
    \multicolumn{1}{|c|}{SVM} & \multicolumn{1}{c|}{0.900} & \multicolumn{1}{c|}{0.866} &
    \multicolumn{1}{c|}{ 0.800} &
    \multicolumn{1}{c|}{ 0.800} \\ \hline
    \multicolumn{1}{|c|}{Logistic Regression} & \multicolumn{1}{c|}{0.750} & \multicolumn{1}{c|}{0.700} &
    \multicolumn{1}{c|}{ 0.650} &
    \multicolumn{1}{c|}{ 0.600} \\ \hline
    \multicolumn{1}{|c|}{Random Forest}           & \multicolumn{1}{c|}{{\bf 0.916}} & \multicolumn{1}{c|}{{\bf 0.900}} &
    \multicolumn{1}{c|}{ 0.800} &
    \multicolumn{1}{c|}{ 0.750} \\ \hline
    \multicolumn{1}{|c|}{Naive Bayes}          & \multicolumn{1}{c|}{0.900} & \multicolumn{1}{c|}{0.866}&
    \multicolumn{1}{c|}{ \bf{0.850}} &
    \multicolumn{1}{c|}{ \bf{0.900}} \\ \hline
    \end{tabular}
        \vspace{-0.1in}
\end{table}

Table \ref{tab:kernelpca-results} shows the results for the classifiers trained using the first two components projections of the RBF Kernel PCA. Unlike the results for linear PCA, the performance of the classifiers is diminished with respect to the results in the original 20-dimensional space. In this case, the best results are reported by SVM, with an accuracy and $F_1$ score of 78.3\% and 76.6\%, respectively.

\begin{table}[h]
    \caption{Performance scores for the classifiers in the 2-D RBF Kernel PCA space.}
    \label{tab:kernelpca-results}
    \centering
    \begin{tabular}{ccccc}
    \hline
    \multicolumn{1}{|c|}{\bf Classifier} &
      \multicolumn{1}{c|}{\bf Accuracy} &
      \multicolumn{1}{c|}{\bf $F_1$ score} &
      \multicolumn{1}{c|}{\bf Precision} &
      \multicolumn{1}{c|}{\bf Recall} \\ \hline
    \multicolumn{1}{|c|}{SVM} & \multicolumn{1}{c|}{{\bf 0.783}} & \multicolumn{1}{c|}{{\bf 0.766}} &
    \multicolumn{1}{c|}{ \bf{0.800}} &
    \multicolumn{1}{c|}{ 0.700} \\ \hline
    \multicolumn{1}{|c|}{Logistic Regression} & \multicolumn{1}{c|}{0.600} & \multicolumn{1}{c|}{0.520}&
    \multicolumn{1}{c|}{ 0.516} &
    \multicolumn{1}{c|}{ \bf{0.800}} \\ \hline
    \multicolumn{1}{|c|}{Random Forest}           & \multicolumn{1}{c|}{0.700} & \multicolumn{1}{c|}{0.700}&
    \multicolumn{1}{c|}{ 0.750} &
    \multicolumn{1}{c|}{ 0.700} \\ \hline
    \multicolumn{1}{|c|}{Naive Bayes}          & \multicolumn{1}{c|}{0.600} & \multicolumn{1}{c|}{0.520}&
    \multicolumn{1}{c|}{ 0.516} &
    \multicolumn{1}{c|}{ 0.800} \\ \hline
    \end{tabular}
        \vspace{-0.1in}
\end{table}

One interesting finding of these results is that the linear PCA representation of the document embeddings yields to better classification results. This is worth noting as the document embeddings often contain an underlying non-linear structure. Because Doc2Vec embeds the document vectors in a semantic vector space, this suggests that the content of the emails can be segmented according to the class through a linear transformation of the vectors.

\iffalse
\begin{table}[h]
    \caption{Accuracy and $F_1$ scores for the classifiers in the 20-dimensional Doc2Vec feature space (3 classes).}
    \label{tab:fulldim-results-3class}
    \centering
    \begin{tabular}{ccc}
    \hline
    \multicolumn{1}{|c|}{\bf Classifier} &
      \multicolumn{1}{c|}{\bf Accuracy} &
      \multicolumn{1}{c|}{\bf $F_1$ score} \\ \hline
    \multicolumn{1}{|c|}{SVM} & \multicolumn{1}{c|}{ 0.291} & \multicolumn{1}{c|}{0.275} \\ \hline
    \multicolumn{1}{|c|}{Logistic Regression} & \multicolumn{1}{c|}{{\bf 0.300}} & \multicolumn{1}{c|}{{\bf 0.279}}\\ \hline
    \multicolumn{1}{|c|}{Random Forest}           & \multicolumn{1}{c|}{0.241} & \multicolumn{1}{c|}{0.233}\\ \hline
    \multicolumn{1}{|c|}{Naive Bayes}          & \multicolumn{1}{c|}{0.233} & \multicolumn{1}{c|}{0.220}\\ \hline
    \end{tabular}
\end{table}
\fi

\subsection{Visualization of Email Embeddings}

\begin{figure}[h]
  \centering
  \includegraphics[width=0.75\linewidth]{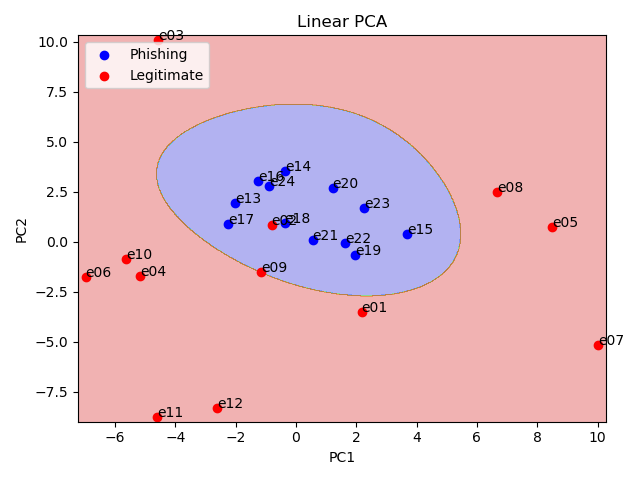}
  \caption{SVM: decision boundary on the 2-D linear PCA projection.}
  \label{fig:svm-pca-bd}
      \vspace{-0.1in}
\end{figure}

Figure \ref{fig:svm-pca-bd} shows the 2-dimensional plot for the email embeddings using the first two linear PCA components where the x-axis represents the first principal and the y-axis holds the values for the second principal. Additionally, we plotted the decision boundary of a SVM with RBF kernel that was fitted using the 24 emails only for visualization purposes. It is visible that the distribution of document embeddings follows a pattern that is easily captured by the decision boundary of the SVM. In this projection, phishing emails are contained in a blob located near the center of the plot, with only two legitimate emails within or in the border of the decision boundary. This finding agrees with the notoriously high performance of the classifiers using the linear PCA 2-D projections. However, note that even though the projection is obtained as linear, the decision boundary is not linear, which also explains the high performance for non-linear classifiers.

\begin{figure}[h]
  \centering
  \includegraphics[width=0.73\linewidth]{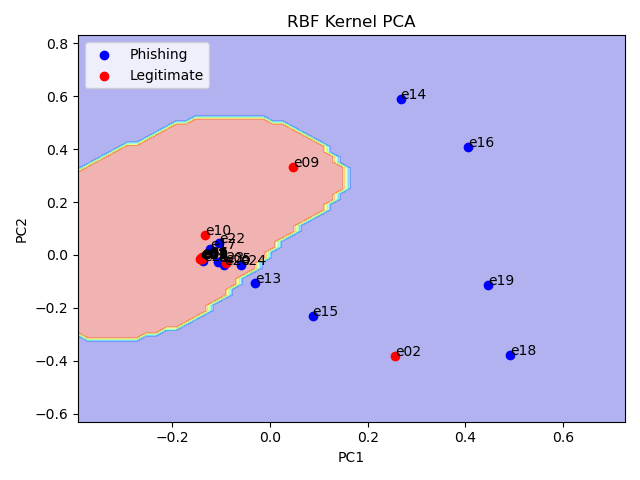}
  \caption{SVM: decision boundary on the 2-D Kernel PCA projection.}
  \label{fig:svm-kernel-pca-bd}
      \vspace{-0.1in}
\end{figure}

Figure \ref{fig:svm-kernel-pca-bd} shows the 2-dimensional plot using the first two RBF Kernel components. Like Figure \ref{fig:svm-pca-bd}, we also plotted the decision boundary of a SVM that was fitted using all the samples and a RBF kernel. However, unlike Figure \ref{fig:svm-pca-bd}, the document embeddings are not clearly segmented by class. Legitimate emails are grouped closely together with the exception of emails 9 and 2, which are the same emails that lie within or near the wrong side of the decision boundary in Figure \ref{fig:svm-pca-bd}. Nevertheless, there are several phishing emails on the wrong side of the decision boundary. In this sense, this projection presents a less useful representation for class segmentation when compared to linear PCA, which can be seen in its classification results in Table \ref{tab:kernelpca-results}. The slight difference between the results in Table \ref{tab:fulldim-results} and Table \ref{tab:kernelpca-results} might suggest that the document embeddings follow a similar configuration in the 20-dimensional feature space.

\subsection{Explained Variance of Linear PCA}

\begin{figure}[h]
  \centering
  \includegraphics[width=0.73\linewidth]{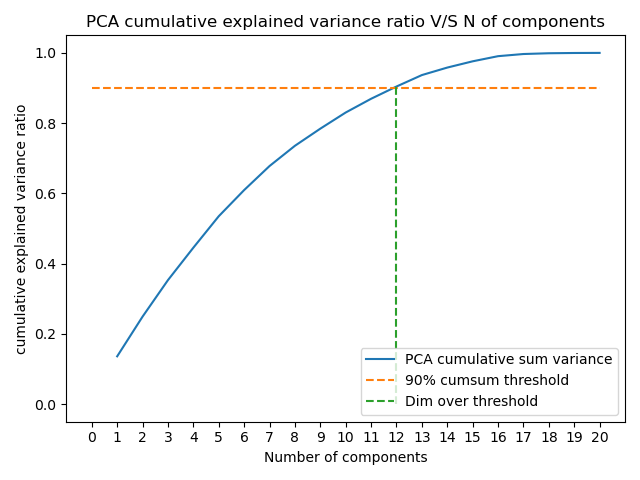}
  \caption{Cumulative explained variance of linear PCA vs. \# of components.}
  \label{fig:linear-pca-cumsum}
      \vspace{-0.1in}
\end{figure}

Given that we obtained better classification results using the linear PCA projection, we performed an explained variance analysis for this PCA. Figure \ref{fig:linear-pca-cumsum} shows the cumulative explained variance per number of components using PCA. The x-axis represents the number of components; whereas, the y-axis shows the cumulative explained variance ratio $R_n$, defined as $R_n = \sum_{i = 1}^n r_i$, where $n$ is the number of components and $r_i$ is the portion of variance explained by the $i$-th component. In this plot, the 90\% of the explained variance is reached using 12 components. The results in Table \ref{tab:linearpca-results} are reported using the first two PCA components, which approximately comprise the 25\% of the total variance. Hence, one surprising finding is that the 91.6\% accuracy of the random forest classifier was achieved using only 25\% of the original variance.

\section{Conclusion and Future Work}
\label{sec:conclusion}

In this work, we generated document embeddings using Doc2Vec and performed binary classification using SVM, Logistic Regression, Random Forest, and Naive Bayes. We executed the experiments on the full 20-dimensional feature space generated by Doc2Vec, and additionally we calculated the 2-D projections of linear PCA and RBF Kernel in order to run the experiments in the low-dimensional projections.

In the original feature space, SVM reports the highest classification performance with an accuracy and $F_1$ score of 81.6\% and 76.6\%, respectively. The results using the 2-D linear PCA projections are higher than those of the 20-dimensional feature space, where the Random Forest reports an accuracy and $F_1$ score of 91.6\% and 90.0\%, respectively, using only 25\% of the original variance. The results using the 2-D RBF Kernel PCA projections are slightly lower than those of the 20-dimensional feature space, where the SVM reports an accuracy and $F_1$ score of 78.3\% and 76.6\%, respectively. The highest results using the linear PCA projections suggests that the underlying structure of the Doc2Vec document embeddings is likely to be linear. The overall high classification results suggest that the semantic vector space in which the document vectors are is appropriate for this classification task. Moreover, the semantics of the emails' content are well suited for the class segmentation.

As future work, we will explore the use of features that permit an easier interpretation and provide a deeper insight into phishing and legitimate emails. There are some other intriguing approaches to address the phishing email detection problem. A possible approach is the use of evidence theory and fusion in formulating the problem \cite{DBLP:conf/bigdataconf/ChatterjeeND18} where a set of linguistic features and evidence can be used to decide pignistic probability of whether an email is phishing. It is also possible to model the phishing email detection through exploring some other machine learning techniques \cite{DBLP:conf/bigdataconf/AbriSKSN19} or emerging deep/machine learning techniques such as reinforcement learning \cite{chatterjee2019detecting}.

\section*{Acknowledgment}
This research work is supported by National Science Foundation (NSF) under Grant No: 1723765.

\bibliographystyle{IEEEtran}
\bibliography{IEEEfull,sample-base}

\end{document}